\documentstyle[pre,aps,eqsecnum,epsf,rotate]{revtex}
\begin{document}
\draft
\tighten
\title{\LARGE Crossover from Isotropic to Directed Percolation}
\vspace{1cm}

\author{Erwin Frey}
\address{Institut f\"ur Theoretische Physik,
Physik-Department der Technischen Universit\"at M\"unchen, \\
James-Franck-Stra\ss e, D-85747 Garching, Germany}

\author{Uwe Claus T\"auber}
\address{Lyman Laboratory of Physics, Harvard University,
Cambridge, Massachusetts 02138, U.S.A.}

\author{Franz Schwabl}
\address{Institut f\"ur Theoretische Physik,
Physik-Department der Technischen Universit\"at M\"unchen, \\
James-Franck-Stra\ss e, D-85747 Garching, Germany}
\date{\today}

\maketitle
\widetext
\centerline{(to be published in Phys. Rev. E, May 1994)}
%\receipt{...... 1993}

%\centerline{( \today )}
\baselineskip = 20pt

\begin{abstract}
Percolation clusters are probably the simplest example for scale--invariant
structures which either are governed by isotropic scaling--laws
(``self--similarity'') or --- as in the case of directed percolation --- may
display anisotropic scaling behavior (``self--affinity''). Taking advantage of
the fact that both isotropic and directed bond percolation (with one preferred
direction) may be mapped onto corresponding variants of (Reggeon) field theory,
we discuss the crossover between self--similar and self--affine scaling. This
has been a long--standing and yet unsolved problem because it is accompanied by
different upper critical dimensions: $d_c^{\rm I} = 6$ for isotropic, and
$d_c^{\rm D} = 5$ for directed percolation, respectively. Using a generalized
subtraction scheme we show that this crossover may nevertheless be treated
consistently within the framework of renormalization group theory. We identify
the corresponding crossover exponent, and calculate effective exponents for
different length scales and the pair correlation function to one--loop order.
Thus we are able to predict at which characteristic anisotropy scale the
crossover should occur. The results are subject to direct tests by both
computer simulations and experiment. We emphasize the broad range of
applicability of the proposed method.
\end{abstract}

\pacs{PACS numbers: 0520, 0540, 6460}

%\pacs{00.00.xx}
%\topmargin= -0.2in
%\baselineskip=20pt

%\newpage

%%%%%%%%%%%%%%%%%%%%%%%%%%%%%%%%%%%%%%%%%%%%%%%%%%%%%%%%%%%%%%%%%%%%%%%%%%%%%%%

\baselineskip = 15pt

\section{Introduction}

In describing a large variety of patterns in nature concepts from fractal
geometry \cite{Man82} have become of increasing importance over the last
years. The simplest kind of scale invariance is {\it self--similarity}, i.e.,
invariance with respect to homogeneous dilation or contraction. A second kind
of structures, which frequently appears in growth models, are {\it
self--affine} clusters, which we define as characterized by {\it anisotropic
scaling} (see Fig. 1).

Perhaps the most simple growth model which incorporates both of these fractal
structures is percolation \cite{Ess80}. In ordinary percolation, sites or bonds
are filled at random with probability $p$. The percolation process then
proceeds along paths connecting occupied nearest neighbors. The clusters formed
by nearest--neighbor links are self--similar, i.e., they display isotropic
scaling. In directed percolation \cite{Kin83} the links between nearest
neighbors have a bias in one preferred direction, such that the percolation
process advances along this direction only. (Often this direction is referred
to as the time direction $t$ and therefore directed percolation proceeds in the
direction of increasing time.) The size of the clusters in the preferred
direction is characterized by a length scale different from that in the
perpendicular direction. Fig. 1 depicts typical clusters emerging from
isotropic (a) and directed (b) percolation, respectively.

If percolation in the positive $t$ direction is merely favored with a certain
probability with respect to the negative $t$ direction, but propagation
``backward'' in time is still admitted, the situation will be more complicated.
If the ``anisotropy'' $g$ is low, one expects almost isotropic scaling behavior
in a large region of the phase diagram. However, when  the critical region
near the percolation threshold $p_c$ is approached, self--affine scaling will
become apparent. Similarly, if $p = p_c$, and the percolation clusters are
viewed at a tiny length scale, hardly any deviations from self--similarity will
be noticeable. On the other hand, if one proceeds to larger and larger scales,
anisotropic scaling will become more and more important, until finally the
asymptotic limit of directed percolation is reached.

Our aim is to provide a quantitative description of the ensuing crossover
features from self--similar to self--affine scaling, when either the length
scale is varied, or the threshold $p_c$ is approached, for a biased percolation
problem. More precisely, we are going to compute at which characteristic
anisotropy scale $g$ this crossover should occur, and we shall find that this
crossover point depends on the specific scaling law and correspondent critical
exponent under consideration. These predictions are, of course, subject to
direct tests by computer simulations and/or experiments, which would thus be
most desirable in order to check our results.

For the issues we have in mind, the main quantity of interest is the pair
correlation function (or connectivity) $G({\bf r}_2,{\bf r}_1)$, which measures
the probability that the sites ${\bf r}_2$ and ${\bf r}_1$ are connected by
some path irrespective of the other sites in the lattice. Lines of constant $G$
hence describe the average shape of the percolating structure. In directed
percolation ${\bf r} = ({\bf x},t)$ the pair correlation function has to be
causal, i.e., one has to add the restriction that $t_2 > t_1$ for $G$ to be
non--zero.

Percolation problems may be reformulated in terms of certain field theories.
Cardy and Sugar \cite{Car80} have shown that directed bond percolation is in
the same universality class as Reggeon field theory, which has been studied
intensively by particle physicists in the 1970s, below five dimensions. This
universality was also confirmed numerically for $d=2$ and $d=3$ (see Refs.
\cite{Bro78,Hen90,Gra89}). A corresponding mapping for isotropic site--bond
correlation onto a related field--theoretical model was performed by Benzoni
and Cardy \cite{Car84}, valid for $d<6$ dimensions. An extension of these
models to the intermediate case of anisotropic, but not entirely directed
percolation is straightforward, and we shall use the ensuing field theory for
our investigation of the crossover from self--similar to self--affine scaling.

The problem of directed percolation has been studied by various theoretical
methods, such as high--temperature expansion for the calculation of the
exponents in $d=2$ \cite{Bro78}, $\epsilon$ expansion \cite{Car80}, and Monte
Carlo simulations \cite{Gra79,Ben84,Gra89}. The main focus of all these
investigations was the determination of the independent critical exponents.

There are many processes in nature which can be described in terms of biased
percolation. An example is the reaction of polymerization with the production
of a giant macromolecule (gelation or vulcanization), if it occurs under
anisotropic external conditions. If the seed macromolecule is washed by a flow
of solution containing monomer groups, it is obvious that the probability of
connection along and against the flow is different \cite{Obu80}. In general,
directed percolation can be understood as a prototypical model for the
spreading of some influence, such as transport in a strong external field
\cite{Lie81}, crack propagation \cite{Ker80}, epidemics or forest fires with a
bias \cite{Gra85}. One more example is the propagation of excitations in the
system of neurons or neuron--like automatons.

Our paper is organized as follows. In the following section we start with an
outline of the mapping of percolation problems onto ``dynamical'' field
theories \cite{Jan76}. We define the specific model under consideration here
and comment on the different upper critical dimensions which play a role in the
limiting cases of isotropic and directed percolation, respectively. In Sec. III
we shall then describe an appropriate renormalization procedure allowing for a
detailed analysis of the crossover scenario. It comprises a generalization of
Amit and Goldschmidt's procedure designed for bicritical points \cite{Ami78}
enabling us to treat the fixed points with different upper critical dimension
within a unified renormalization scheme. The fourth section is devoted to the
solution of the renormalization group equation for the two--point vertex
function and the discussion of the resulting flow equations (including their
scaling behavior) for the coupling parameters in the framework of an explicit
one--loop theory. On this basis, we shall identify the asymptotic critical
indices for both isotropic and directed percolation, and an additional
crossover exponent $\Delta$. Finally, we shall calculate effective critical
exponents for the different length scales and the wavevector dependence of the
pair correlation function, which allows us to determine the relevant crossover
scales. In the Appendices, we list some technical details concerning the
one--loop perturbation theory, and some properties of those integrals that
emerge after the application of Feynman's parametrization, and enter the
explicit expressions for the renormalization constants and flow equations.

%%%%%%%%%%%%%%%%%%%%%%%%%%%%%%%%%%%%%%%%%%%%%%%%%%%%%%%%%%%%%%%%%%%%%%%%%%%%%%%

\section{Mapping to a Field--theoretical Model}

Following the considerations by Cardy and Sugar \cite{Car80}, and Benzoni and
Cardy \cite{Car84}, we argue that our general percolation problem can be mapped
onto a field--theoretical model. For the reader's convenience, a very brief
sketch of the basic ideas entering the derivation of the probability measure is
presented here; more details may be found in the literature
\cite{Ess80,Car80,Car84}. As stated above, the central quantity of interest is
the pair--correlation function $G^0({\bf r}_2,{\bf r}_1)$ (we shall henceforth
denote unrenormalized quantities with a superscript ``$0$''), or probability
that sites ${\bf r}_1 = ({\bf x}_1,t_1)$ and ${\bf r}_2 = ({\bf x}_2,t_2)$
belong to the same percolation cluster. In a first step,
$G^0({\bf r}_2,{\bf r}_1)$ is represented by a sum of all such graphs defined
on the percolating lattice which are constructed by the following rules
\cite{Ess80,Car80,Car84} (see Fig. 2):

(1) Place oriented bonds on the lattice, in such a fashion that from each site
in the diagram it is possible to reach ${\bf r}_2$ by following the arrows
forward, and ${\bf r}_1$ by following the arrows backward.
(2) Closed loops of arrows are not allowed.
(3) Insert a factor $p$ for each bond.
(4) In order to avoid multiple countings, insert a factor $-i^n$ for each
vertex where $n$ bonds meet.

The second step is a more formal expression for $G^0({\bf r}_2,{\bf r}_1)$, to
be obtained by introducing commuting ``ladder'' operators $a({\bf r}_i)$ and
${\tilde a}({\bf r}_i)$ at each site ${\bf r}_i$, and an operation $Tr$, with
the properties
\begin{equation}
     a^2 = i a \quad , \qquad {\tilde a}^2 = i {\tilde a} \quad ,
\label{2.1}
\end{equation}
\begin{equation}
 Tr \, a = Tr \, {\tilde a} = 0 \quad , \qquad Tr ( a {\tilde a} ) = 1 \quad .
\label{2.2}
\end{equation}
Denoting by $P$ an operator projecting out the graphs with closed loops, the
diagrammatic rules above can now all be summarized in the formula
\begin{equation}
     G^0({\bf r}_2,{\bf r}_1) = Tr \left( P \, a({\bf r}_1)
\prod_{\langle links \rangle ; i,j} [ 1 + p {\tilde a}({\bf r}_i) a({\bf r}_j)]
                                                a({\bf r}_2) \right) \quad .
\label{2.3}
\end{equation}

Introducing
\begin{equation}
     \lambda = - \ln (1-p) \quad ,
\label{2.4}
\end{equation}
the ``transition probability'' in Eq.~(\ref{2.3}) may be exponentiated
according to
\begin{equation}
  \prod_{\langle links \rangle ; i,j}
   [ 1 + p {\tilde a}({\bf r}_i) a({\bf r}_j)]
        = \prod_{\langle links \rangle ; i,j}
              \exp \left[ \lambda {\tilde a}({\bf r}_i) a({\bf r}_j) \right] =
\exp \left[ \sum_{ij} {\tilde a}({\bf r}_i) V_{ij} a({\bf r}_j) \right] \quad ;
\label{2.5}
\end{equation}
in the final expression here, $V_{ij} = v({\bf r}_i - {\bf r}_j)$ is a matrix
which stems from the nearest--neighbor interaction $\lambda$ and depends on the
details of the lattice structure, and $v({\bf r})$ is a short--ranged
function. The next step in our derivation is a Gaussian transformation, i.e., a
representation of Eq.~(\ref{2.3}) by a functional integral over auxiliary
fields $\phi_0$ and ${\tilde \phi}_0$:
\begin{eqnarray}
    G^0({\bf r}_2,{\bf r}_1) = Tr \Biggl\{ P \, a({\bf r}_1)
      \int {\cal D}[\phi_0] \int {\cal D}[{\tilde \phi}_0] \,
       &&\exp \left( - \sum_{i,j} {\tilde \phi}_0({\bf r}_i) V_{ij}^{-1}
                                 \phi_0({\bf r}_j) \right) \times \label{2.6}\\
  &&\times \exp \left( \sum_i \left[ {\tilde a}({\bf r}_i) \phi_0({\bf r}_i) +
                        a({\bf r}_i) {\tilde \phi}_0({\bf r}_i) \right] \right)
                                       a({\bf r}_2) \Biggr\} \quad . \nonumber
\end{eqnarray}

Now we take the continuum limit, and expand $V_{ij}^{-1}$ with respect to
gradients of ${\bf r}$; to lowest order, one has
\begin{equation}
     v^{-1} = {1 \over n_c} \left( 1 - s {\bf \nabla}_{{\bf x}}^2
                                      + r_1 {\partial \over \partial t}
              - r_2 {\partial^2 \over \partial t^2} + \ldots \right) \quad ,
\label{2.7}
\end{equation}
where $n_c$ is the coordination number of the lattice. Note that we have
explicitly taken account of the fact that our problem displays an inversion
symmetry with respect to ${\bf x}$; in the case of isotropic percolation the
term $\propto r_1$ vanishes, and (\ref{2.7}) is even invariant under
transformations $t \rightarrow -t$, and hence ${\bf r} \rightarrow -{\bf r}$,
while in the self--affine region the second ``time'' derivative becomes
irrelevant.

At last, we have to perform the operation $Tr$ in Eq.~(\ref{2.6}); this may be
achieved by expanding the exponential with respect to powers of $a({\bf r}_i)$
and ${\tilde a}({\bf r}_i)$. For the detailed calculations, we refer the
interested reader to the Appendix 1 of Ref. \cite{Car84}. The final result for
the pair correlation function is a sum over $(m+n)$ point correlation functions
\begin{equation}
     G^0({\bf x}_2,t_2;{\bf x}_1,t_1) =
      \sum_{m,n=1}^\infty {(-i)^{m+n-2} \over m! n!}
      G^0_{mn} ({\bf x}_2,t_2;{\bf x}_1,t_1) \quad ;
\label{2.8}
\end{equation}
here the $G^0_{mn}({\bf x}_2,t_2;{\bf x}_1,t_1)$ are defined via
\begin{eqnarray}
      G^0_{mn} ({\bf x}_2,t_2;{\bf x}_1,t_1) &&=
      \langle \phi_0({\bf x}_2,t_2)^m {\tilde \phi}_0({\bf x}_1,t_1)^n \rangle
                                                              \nonumber \\
      &&= P \int {\cal D} [\phi_0] \int {\cal D} [{\tilde \phi}_0] \;
      \phi_0({\bf x}_2,t_2)^m {\tilde \phi}_0({\bf x}_1,t_1)^n
      \exp \left[ {\cal J}[\phi_0,{\tilde \phi}_0] \right] \quad , \label{2.9}
\end{eqnarray}
where the probability measure explicitly reads
\begin{eqnarray}
     {\cal J}[\phi_0,{\tilde \phi}_0] = - \int d^Dx \int dt
      \biggl\{ &&{\tilde \phi}_0({\bf x},t)
      \left[ r_0 - \nabla^2 - {1 \over c_0^2} {\partial^2 \over \partial t^2}
             + {2 g_0 \over c_0} {\partial \over \partial t}
      \right] \phi_0({\bf x},t) \nonumber \\
     &&+ {u_0 \over 2} \left[ {\tilde \phi}_0({\bf x},t)^2 \phi_0({\bf x},t) -
     {\tilde \phi}_0({\bf x},t) \phi_0({\bf x},t)^2 \right] \biggr\} \quad .
\label{2.10}
\end{eqnarray}
When writing down Eq.~(\ref{2.10}), we have omitted nonlinear terms of higher
than third order with respect to the fields $\phi_0$ and ${\tilde \phi}_0$ for
universality (we are interested in the scaling behavior near the percolation
threshold $p_c$) and renormalizability reasons; the neglected contributions
constitute irrelevant perturbations in the sense of the renormalization group.
Furthermore the expansion parameters of Eq.~(\ref{2.7}) have been renamed for
convenience. For the percolation threshold itself one finds
\begin{equation}
     p_c = 1 - e^{-1 / n_c} \quad ,
\label{2.11}
\end{equation}
and for $p \rightarrow p_c$ one has
\begin{equation}
    r_0 \propto p - p_c \quad .
\label{2.12}
\end{equation}
In Eq.~(\ref{2.9}), the operator $P$ now assures ``causality'', and projects
out ``acausal'' diagrams, e.g.: closed Hartree loops containing the
``response'' propagator (compare Ref. \cite{Jan76}).

On the basis of Eqs.~(\ref{2.8} -- \ref{2.10}) one is now in a position to
construct a perturbation expansion for the pair correlation function
$G^0({\bf r}_2,{\bf r}_1)$. In this paper, however, we shall refine ourselves
to the study of the renormalization group equation for the two--point vertex
function
\begin{equation}
 \Gamma^0_{11} ({\bf q},\omega) = {1 \over G^0_{11}(-{\bf q},-\omega)} \quad ,
\label{2.13}
\end{equation}
which already displays the correct scaling behavior and will permit the
identification of the relevant critical exponents. Here we have defined the ($d
= D + 1$) dimensional Fourier transformation according to
\begin{equation}
    \phi_0({\bf x},t) = \int_q \int_\omega \phi_0({\bf q},\omega)
                                    e^{i ({\bf q} {\bf x} - \omega t)} \quad ,
\label{2.14}
\end{equation}
where we have introduced the convenient abbreviation
\begin{equation}
    \int_q \int_\omega \ldots =
    {1 \over (2 \pi)^d} \int d^Dq \int d\omega \ldots \quad .
\label{2.15}
\end{equation}
For the renormalization of the nonlinear coupling $u_0$, we shall also have to
consider the two three--point vertex functions
\begin{eqnarray}
     \Gamma^0_{12} \left( -{\bf q},-\omega;{{\bf q} \over 2},{\omega \over 2};
                                 {{\bf q} \over 2},{\omega \over 2} \right) &&=
      - {G^0_{21} \left( -{{\bf q} \over 2},-{\omega \over 2};
                    -{{\bf q} \over 2},-{\omega \over 2};{\bf q},\omega \right)
 \over G^0_{11}({\bf q},\omega) G^0_{11}({{\bf q} \over 2},{\omega \over 2})^2}
                                    \quad ,                  \label{2.16}\\
      \Gamma^0_{21} \left( {{\bf q} \over 2},{\omega \over 2};
                {{\bf q} \over 2},{\omega \over 2};-{\bf q},-\omega \right) &&=
      - {G^0_{12} \left( {\bf q},\omega;-{{\bf q} \over 2},-{\omega \over 2};
                                  -{{\bf q} \over 2},-{\omega \over 2} \right)
\over G^0_{11}({\bf q},\omega) G^0_{11}({{\bf q} \over 2},{\omega \over 2})^2}
                                                          \quad . \label{2.17}
\end{eqnarray}
The practical advantage of using these vertex functions instead of the
correlation functions themselves is their correspondence to the one--particle
irreducible diagrams within the graphical representation in terms of Feynman
diagrams (see e.g. Ref. \cite{Ami84}). (We remark that in the definitions of
the vertex functions (\ref{2.13}), (\ref{2.16}), and (\ref{2.17}), the Dirac
$\delta$ functions stemming from translational symmetry in ${\bf x}$ and $t$
have been split off.)

Returning to the ``dynamical'' functional \cite{Jan76} (\ref{2.10}), it is
easily seen that setting the parameter $g_0$ to zero leads to the
field--theoretical model of Benzoni and Cardy for the special case of isotropic
bond percolation in $d = D + 1$ dimensions \cite{Car84}. Of course, $c_0$ may
then be assumed to take the value $1$. On the other hand, in the limit $g_0
\rightarrow \infty$ and $c_0 \rightarrow \infty$ such that $g_0 / c_0$ remains
finite, one obtains Cardy and Sugar's Reggeon field theory for the problem of
directed percolation with $t$ denoting the preferred direction \cite{Car80}.
Any finite value of $g_0$ hence corresponds to a biased percolation problem,
with $g_0$ characterizing the strength of the inherent anisotropy.

We now proceed with a simple dimensional analysis to determine the upper
critical dimensions of our model in the different limiting cases. If we define
\begin{equation}
    [ x ] = [ t ] = \Lambda^{-1} \quad ,
\label{2.18}
\end{equation}
where $\Lambda$ is a cutoff wave vector which defines the microscopic length
scale of the problem, then we find for the primitive dimension of the
stochastic fields (using $[ J ] = \Lambda^0$)
\begin{equation}
   [ \phi_0({\bf x},t) ] = [ {\tilde \phi}_0({\bf x},t) ] = \Lambda^{(2 - d)/2}
     \quad .
\label{2.19}
\end{equation}
Hence the coupling parameters acquire the following ``naive'' dimensions
\begin{equation}
      \left[ r_0 \right] = \Lambda^2   \, , \quad
      \left[ c_0 \right] = \Lambda^0   \, , \quad
      \left[ g_0 \right] = \Lambda^1   \, , \quad {\rm and} \quad
      \left[ u_0 \right] = \Lambda^{(d - 6)/2} \quad .
\label{2.20}
\end{equation}

The upper critical dimension may be identified by noting that the relevant
nonlinear coupling has zero primitive dimension at $d = d_c$. In the isotropic
case ($g_0 = 0$), the expansion parameter of the perturbation series turns out
to be $u_0^2 c_0$, and hence the upper critical dimension for isotropic
percolation is found to be $d_c^{\rm I} = 6$. In the extremely anisotropic
limit, $g_0 \rightarrow \infty$ (and $g_0 / c_0$ = const.), on the other hand,
a simple rescaling of the fields shows that now the effective coupling is
$u_0^2 c_0 / g_0$, and using Eq.~(\ref{2.20}) demonstrates that $d_c^{\rm D} =
5$ for directed percolation. Hence accompanying the crossover from
self--similar to self--affine scaling, there is a change of the upper critical
dimension. At first sight, this renders this crossover problem rather
cumbersome, at least if one wants to use an ($\epsilon = d_c - d$) expansion
near $d_c$. We shall see, however, that a procedure similar to Amit and
Goldschmidt's treatment of bicritical points \cite{Ami78} will enable us to
give a consistent mathematical description of this crossover. Of course, we
shall have to refrain from any $\epsilon$ expansion. We would like to remark
that different modifications of this (somewhat misleadingly) so--called
``generalized minimal subtraction scheme'' have been successfully employed by
Lawrie \cite{Law81} and the present authors \cite{Fre88,Tau92,Tau93} for
further interesting crossover scenarios.

%%%%%%%%%%%%%%%%%%%%%%%%%%%%%%%%%%%%%%%%%%%%%%%%%%%%%%%%%%%%%%%%%%%%%%%%%%%%%%%

\section{Perturbation Theory and Renormalization}

On the basis of Eq.~(\ref{2.10}), one may derive the perturbation series with
respect to $u_0$ following the common procedure (see e.g. Ref. \cite{Ami84}).
{}From the bilinear part of the ``dynamic''functional ${\cal J} [\phi_0,
{\tilde
\phi}_0]$, one easily derives the free propagator
\begin{equation}
     G^0_{11 \, (0)}({\bf q},\omega) =
     {1 \over r_0 + q^2 + \omega^2 / c_0^2 - 2 i \omega g_0 / c_0} \quad ,
\label{3.1}
\end{equation}
and the vertices may be read off from the anharmonic part of (\ref{2.10}). In
Fig. 3, we depict these elements for constructing the Feynman graphs
of our field--theoretical representation of the biased percolation problem.

For $r_0 \rightarrow 0$, the ensuing perturbation theory is of course
infrared--divergent, leading to non--trivial critical exponents. These
anomalous dimensions are derived via studying the ultraviolet singularities of
the field theory, which appear at the upper critical dimension $d_c$, when the
momentum cutoff $\Lambda$ is pushed to infinity (see e.g. Ref. \cite{Ami84}).
Finite values are then assigned to these UV--divergent integrals through the
application of a regularization prescription. We shall choose the dimensional
regularization scheme as introduced by t'Hooft and Veltman \cite{Hoo72}; here
the ($\Lambda \rightarrow \infty$) singularities appear as poles $\propto 1 /
(d_c-d)$.

The ultraviolet divergences may then be collected in renormalization constants
and absorbed into the definition of multiplicatively renormalized quantities.
Thus we define the renormalized fields
\begin{equation}
             \phi  = Z_\phi^{1/2}         \phi_0 \quad , \qquad
      {\tilde \phi} = Z_\phi^{1/2} {\tilde \phi}_0 \quad ,
\label{3.2}
\end{equation}
and the renormalized parameters
\begin{eqnarray}
    r   &&= Z_\phi^{-1}   Z_r (r_0 - r_{0c}) \mu^{-2} \quad , \label{3.3}\\
     c^2 &&= Z_\phi        Z_c^{-1}     c_0^2        \quad ,   \label{3.4}\\
     g   &&= Z_\phi^{-1/2} Z_c^{-1/2} Z_g g_0 \mu^{-1} \quad , \label{3.5}\\
     u   &&= Z_\phi^{-3/2} Z_u u_0 B_d^{1/2}  \mu^{(d-6)/2} \quad .
                                                               \label{3.6}
\end{eqnarray}
Note that both fields are renormalized with the same Z factor, implying
that $\Gamma_{11} = Z_\phi^{-1} \Gamma^0_{11}$, $\Gamma_{12} = Z_\phi^{-3/2}
\Gamma^0_{12}$, and $\Gamma_{21} = Z_\phi^{-3/2} \Gamma^0_{21}$.

In (\ref{3.3}) we have taken into account the fact that the fluctuations will
also shift the percolation threshold. Furthermore we have rendered the
renormalized quantities dimensionless by introducing the explicit arbitrary
length scale $1/\mu$, and have finally included the geometric factor
\begin{equation}
     B_d = {\Gamma(4 - d/2) \over (4 \pi)^{d/2}}
\label{3.7}
\end{equation}
in the definition of the renormalized coupling (\ref{3.6}).

In the case of crossover phenomenona, however, there is (at least) one
additional relevant length scale besides the correlation length, given by an
anisotropy or ``mass'' parameter describing the variation from one scaling
region to the other. This implies the technical difficulty that both the UV and
IR singularities will differ in the two distinct scaling regimes. In the
``traditional'' approach to crossover problems, one would compute the critical
exponents in the vicinity of one of the stable fixed points; all the crossover
features would then be contained in the accompanying scaling function as
corrections to this scaling behavior. However, in general a calculation to high
order in the perturbation expansion would be required in order to achieve a
satisfactory description of the entire crossover region. Of course, using an
($\epsilon = d_c - d$) expansion with respect to either of the fixed points
renders the other one completely inaccessible, if their upper critical
dimensions do not coincide. Therefore Amit and Goldschmidt's idea to
incorporate the crossover features already in the exponent functions has proven
much more successful than treating the problem on the basis of scaling
functions. The essential prescription one has to bear in mind is that the
renormalization constants are not solely functions of the anharmonic coupling,
but necessarily also of the additional ``mass'' or anisotropy parameter
describing the interplay between the two different scaling regimes
\cite{Ami78}.

In our case this second length scale is related to the anisotropy parameter
$g_0$. For a consistent treatment of the entire crossover region, one thus has
to assure that the UV singularities are absorbed into the Z factors for any
arbitrary value of $g_0$, including $g_0 \rightarrow \infty$. This is not a
trivial prescription, as usually the $1 / (d_c-d)$ poles will be altered in the
different scaling regimes. For the situation that we have in mind, even the
value of the upper critical dimension is bound to change as the crossover takes
place, in contrast to previously studied cases \cite{Law81,Fre88,Tau92,Tau93}.
However, we shall demonstrate that with the above stated so-called
``generalized subtraction scheme'' this change in the upper critical dimension
may be incorporated into the usual formalism without any drastic changes, if
one refrains from any $\epsilon$ expansion about $d_c$. The
perturbation series is then an expansion with respect to the effective coupling
$v$ to be introduced later, which is not an a-priori small parameter, and the
perturbation expansion is uncontrolled in this sense \cite{Eps}. If higher
orders of the perturbation expansion were known, one could substantially
refine the theory by a Borel resummation procedure. For a more detailed
discussion of the question in which cases one may dispense with a $(d_c-d)$
expansion, we refer to work of Schloms and Dohm \cite{Doh89}.

Another (minor) prize to be paid is that for the flow equations and related
quantities only numerical solutions are accessible, and merely the limiting
cases of $g_0 = 0$ and $g_0 \rightarrow \infty$, respectively, allow for an
analytical investigation. We remark that the somewhat misleading term
``generalized subtraction scheme'' stems from the fact that in the framework of
an $\epsilon$ expansion this corresponds to adding logarithms of $g_0$, which
are finite in the limit $\epsilon \rightarrow 0$, to the $Z$ factors; see the
original work by Amit and Goldschmidt \cite{Ami78}. It should be emphasized
that the procedure described here is obviously applicable to a great variety of
crossover problems (for some examples see Sec. V.).

After these general statements, let us return to our explicit
calculations. In Fig. 4 we have depicted the one--loop diagrams for
$\Gamma^0_{11}$, $\Gamma^0_{12}$, and $\Gamma^0_{21}$. From $\partial_{q^2}
\Gamma_{11}({\bf q},0) \mid_{q = 0}$ one infers directly the field
renormalization $Z_\phi$, while  $Z_r$, $Z_c$, and $Z_g$ can be calculated by
investigation of $\Gamma_{11}({\rm 0},0)$, $\partial_{\omega^2}
\Gamma_{11}({\rm 0},\omega) \mid_{\omega = 0}$, and $\partial_{\omega}
\Gamma_{11}({\rm 0},\omega) \mid_{\omega = 0}$, respectively. Finally $Z_u$ is
to be obtained from one of the renormalized three--point vertex functions at
vanishing external momenta and frequencies (see Appendix A). All these vertex
functions are investigated at finite ``mass'' $r_0 = \mu^2$, in order to avoid
complications stemming from additional infrared singularities. Thus the
renormalization scale $\mu$ comes into play.

The explicit one--loop results for the renormalization factors --- being
functions of the anisotropy scale $g_0$ ! --- read
\begin{eqnarray}
     Z_\phi &= &1 - {u_0^2 c_0 B_d \mu^{d-6} \over 8 (6-d)}
                 \left[ I^d_{15}(g_0 / \mu) - I^d_{33}(g_0 / \mu) \right]
                                                         \quad , \label{3.8}\\
         Z_r &= &1 - {u_0^2 c_0 B_d \mu^{d-6} \over 2 (6-d)}
                        I^d_{15}(g_0 / \mu)              \quad , \label{3.9}\\
         Z_c &= &1 - {u_0^2 c_0 B_d \mu^{d-6} \over 8 (6-d)}
                 \left[ I^d_{15}(g_0 / \mu) - I^d_{33}(g_0 / \mu) \right]
                                                                   \nonumber\\
             &\phantom{=} &\quad + {u_0^2 c_0 g_0^2 B_d \mu^{d-8} \over 8}
                 \left[ 2 I^d_{35}(g_0 / \mu) - I^d_{53}(g_0 / \mu) \right]
                                                         \quad , \label{3.10}\\
         Z_g &= &1 - {u_0^2 c_0 B_d \mu^{d-6} \over 4 (6-d)}
                 \left[ I^d_{15}(g_0 / \mu) - I^d_{33}(g_0 / \mu) \right]
                                                         \quad , \label{3.11}\\
         Z_u &= &1 - {u_0^2 c_0 B_d \mu^{d-6} \over 6-d}
                        I^d_{15}(g_0 / \mu)              \quad , \label{3.12}
\end{eqnarray}
where we have introduced the abbreviations $I^d_{mn}(g_0 / \mu)$ for a certain
class of Feynman parameter integrals which appear in the course of the
calculations (compare Appendix B).

At this point, we may study the behavior of the $Z$ factors in the two limiting
cases of isotropic and directed percolation, respectively. Let us investigate
$Z_u$, for example; for $g_0 = 0$ we find with Eq.~(\ref{B3})
\begin{equation}
     g_0 = 0 \; : \qquad Z_u =
                 1 - {2 u_0^2 c_0 B_d \mu^{d-6} \over 6 - d} \quad ,
\label{3.13}
\end{equation}
while in the opposite case $g_0 \rightarrow \infty$ the pole at $d = 6$ is
cancelled and replaced by another one at $d = 5$
\begin{equation}
      g_0 \rightarrow \infty \; : \qquad Z_u =
      1 - {u_0^2 c_0 B_d B(1/2,(7-d)/2) \mu^{d-5} \over g_0 (5 - d)}
\label{3.14}
\end{equation}
[here Eq.~(\ref{B4}) has been used]. Hence our prescription provides the
relevant effective coupling constants and the correct pole structure in both
limits, and interpolates smoothly in between.

Finally the fluctuation--induced shift of the percolation threshold results as
the solution of the implicit equation
\begin{equation}
     r_{0c} =
      \left[ {u_0^2 c_0 B_d \over (d-4) (6-d)} I^d_{13}(g_0/\sqrt{r_{0c}})
                                            \right]^{2 / (6 - d)} \quad .
\label{3.15}
\end{equation}
Note that $r_{0c}$ is a non-analytical function of $u_0$ for both limits $g_0 =
0$ and $g_0 \rightarrow \infty$ (see e.g. Ref. \cite{Doh89}).

%%%%%%%%%%%%%%%%%%%%%%%%%%%%%%%%%%%%%%%%%%%%%%%%%%%%%%%%%%%%%%%%%%%%%%%%%%%%%%%

\section{Renormalization Group and Flow Equations}

\subsection{Scaling Behavior and Critical Exponents}

The renormalization group equation serves to connect the asymptotic theory,
where the infrared singularities manifest themselves, with a region in
parameter space, where the coupling $u$ is finite (but not necessarily small)
and ordinary ``naive'' perturbation expansion becomes applicable. It explicitly
takes advantage of the scale invariance of the system near a critical point
(i.e., the percolation threshold in our case). More precisely, we observe that
the bare two--point vertex function is of course independent of the arbitrary
renormalization scale $\mu$:
\begin{equation}
      \mu {d \over d \mu} \bigg \vert_0
      \Gamma^0_{11}(r_0,c_0,g_0,u_0,{\bf q},\omega) = 0 \quad .
\label{4.1}
\end{equation}
Introducing Wilson's flow functions
\begin{eqnarray}
     \zeta_\phi = \mu {\partial \over \partial \mu} \bigg \vert_0
                    \ln Z_\phi                          \quad &&, \label{4.2}\\
     \zeta_r = \mu {\partial \over \partial \mu} \bigg \vert_0
                 \ln {r \over r_0 - r_{0c}} &&= - 2 - \zeta_\phi
                  + \mu {\partial \over \partial \mu} \bigg \vert_0 \ln Z_r
                                                          \quad , \label{4.3}\\
     \zeta_c = \mu {\partial \over \partial \mu} \bigg \vert_0
                 \ln {c \over c_0} &&= {1 \over 2} \zeta_\phi
                 - {1 \over 2} \mu {\partial \over \partial \mu} \bigg \vert_0
                                                  \ln Z_c \quad , \label{4.4}\\
     \zeta_g = \mu {\partial \over \partial \mu} \bigg \vert_0
                 \ln {g \over g_0} &&= - 1 - \zeta_\phi + \zeta_c
                  + \mu {\partial \over \partial \mu} \bigg \vert_0 \ln Z_g
                                                          \quad , \label{4.5}\\
     \zeta_u = \mu {\partial \over \partial \mu} \bigg \vert_0
                 \ln {u \over u_0} &&= {d - 6 \over 2}
    - {3 \over 2} \zeta_\phi + \mu {\partial \over \partial \mu} \bigg \vert_0
                                                  \ln Z_u \quad , \label{4.6}
\end{eqnarray}
we may transform Eq.~(\ref{4.1}) into a partial differential equation for the
renormalized vertex function
\begin{equation}
     \left[     \mu {\partial \over \partial \mu} +
             \zeta_r r {\partial \over \partial r} +
             \zeta_c c {\partial \over \partial c} +
             \zeta_g g {\partial \over \partial g} +
             \zeta_u u {\partial \over \partial u} + \zeta_\phi \right]
                           \Gamma_{11}(\mu,r,c,g,u,{\bf q},\omega) = 0 \quad .
\label{4.7}
\end{equation}
The symbol $\vert_0$ indicates that all the derivatives are to be taken at
fixed bare parameters $r_0$, $g_0$, and $u_0$. One should note that, as a
consequence of the generalized renormalization scheme, all the flow functions
$\zeta$ are functions of $u$, $c$, {\it and} $g$.

The renormalization group equation (\ref{4.7}) is now readily solved with the
method of characteristics. The characteristics $a(\ell)$ of Eq.~(\ref{4.7})
define the running parameters and coupling constants into which these transform
when $\mu \rightarrow \mu(\ell) = \mu \ell$. They are given by the solutions to
first--order differential equations ($a=r,c,g,u$)
\begin{equation}
     \ell {d a(\ell) \over d \ell} = \zeta_a(\ell) a(\ell) \quad,
\label{4.8}
\end{equation}
with the initial conditions $r(1) = r$, $c(1) = c$, $g(1) = g$, and $u(1) = u$,
namely
\begin{equation}
     a(\ell) = a e^{\int_1^\ell \zeta_a(\ell^\prime) d \ell'/ \ell'} \quad .
\label{4.9}
\end{equation}
Defining the dimensionless vertex function ${\hat \Gamma}_{11}$ according to
\begin{equation}
     \Gamma_{11}(\mu,r,c,g,u,{\bf q},\omega) =
       \mu^2 {\hat \Gamma}_{11} \left( r,v,{{\bf q} \over \mu},
            {g \omega \over c \mu},{\omega^2 \over c^2 \mu^2} \right) \quad ,
\label{4.10}
\end{equation}
the solution of (\ref{4.7}) reads
\begin{equation}
     \Gamma_{11}(\mu,r,c,g,u,{\bf q},\omega) =
      \mu^2 \ell^2 e^{ \int_1^\ell \zeta_\phi(\ell') d \ell' / \ell'}
      \Gamma_{11} \left( r(\ell), v(\ell), {{\bf q} \over \mu \ell},
                     {g(\ell) \omega \over c(\ell) \mu \ell},
                     {\omega^2 \over c(\ell)^2 \mu^2 \ell^2} \right) \quad .
\label{4.12}
\end{equation}
Here we have introduced an effective anharmonic coupling
\begin{equation}
    v = u^2 c I^d_{17}(g) \quad ,
\label{4.13}
\end{equation}
which acquires finite values in both limits, $g \rightarrow 0$ and $g
\rightarrow \infty$. Defining the corresponding $\beta$ function
\begin{equation}
     \beta_v = \mu {\partial \over \partial \mu} \bigg \vert_0 v \quad ,
\label{4.14}
\end{equation}
the flow of the running coupling $v(\ell)$ is given by the differential
equation
\begin{equation}
     \ell {d v(\ell) \over d \ell} = \beta_v(\ell) \quad .
\label{4.15}
\end{equation}

In the flow equations above, the parameter $\ell$ may be considered as
describing the effect of a scaling transformation upon the system. Obviously,
the theory becomes scale--invariant when a fixed point $v^*$, to be obtained as
a zero of the $\beta$ function,
\begin{equation}
     \beta_v(v^*) = 0 \quad ,
\label{4.16}
\end{equation}
is approached. The properties of $\Gamma_{11}$ in the vicinity of the fixed
point will yield the correct asymptotic behavior, if the latter is
infrared--stable, i.e.: if $\partial  \beta_v / \partial v \big \vert_{v = v^*}
> 0 $ is satisfied, for in this case the flow of the running coupling will
reach $v^*$ for $\ell \rightarrow 0$.

We now turn to the investigation of (\ref{4.12}) near a fixed point $v^*$;
introducing the fixed--point values of the $\zeta$ functions $\zeta_a^* =
\zeta_a(v = v^*)$, also called the anomalous dimensions of the parameters $a$,
we find that $\Gamma_{11}$ becomes a generalized homogeneous function
\begin{equation}
     \Gamma_{11}(\mu,r,c,g,u,{\bf q},\omega) = \mu^2 \ell^{2 + \zeta_\phi^*}
   {\hat \Gamma}_{11} \left( r \ell^{\zeta_r^*}, v^*, {{\bf q} \over \mu \ell},
                  {g \omega \over c \mu \ell^{1 + \zeta_c^* - \zeta_g^*}},
         {\omega^2 \over c^2 \mu^2 \ell^{2 (1 + \zeta_c^*)}} \right) \quad .
\label{4.18}
\end{equation}
Using the matching condition $\ell = q / \mu$, one arrives at the following
general {\it self--affine} scaling form
\begin{equation}
     \Gamma_{11}(\mu,r,c,g,u,{\bf q},\omega) \propto q^{2 - \eta_\perp}
      {\hat \Gamma}_{11} \left( {r \over (q / \mu)^{1 / \nu_\perp}}, v^*, 1,
                 {g \omega \over c \mu (q / \mu)^z},
                {\omega^2 \over c^2 \mu^2 (q / \mu)^{2 z (1 - \Delta)}} \right)
                                                                   \quad ,
\label{4.20}
\end{equation}
where we have defined {\it four} independent critical exponents according to
\begin{equation}
      \eta_\perp = - \zeta_\phi^*    \, ,  \quad
      \nu_\perp = - {1 \over \zeta_r^*}       \, , \quad
              z = 1 + \zeta_c^* - \zeta_g^*   \, , \quad {\rm and } \quad
       z \Delta = - \zeta_g^*                       \quad .
\label{4.21}
\end{equation}
$\eta_\perp$ and $\nu_\perp$ correspond to the two independent indices familiar
from the theory of static critical phenomena. The exponent $z$ was introduced
in analogy to a dynamical critical exponent, and is in our case related to the
anisotropic scaling behavior \cite{Exp}. Finally, $\Delta$ is a crossover
exponent, $0 < \Delta < 1$, describing the transition from isotropic to
directed percolation. It stems from the fact that there appear {\it two}
different scaling variables for the ``frequency'' $\omega$ in Eq.~(\ref{4.20}).
In the asymptotic limit of directed percolation, $g \rightarrow \infty$
(accompanied by $c \rightarrow \infty$ while $g/c$ remains finite), the second
scaling variable vanishes, and the scaling behavior is described by the {\it
three} exponents $\eta_\perp$, $\nu_\perp$, and $z$.

Similarly, with the choice $\ell = (g \omega / c \mu)^{1 / (1 + \zeta_c^* -
\zeta_g^*)}$ Eq. \ref{4.18} reads
\begin{equation}
    \Gamma_{11}(\mu,r,c,g,u,{\bf q},\omega) \propto \omega^{2 - \eta_\parallel}
      {\hat \Gamma}_{11} \left( {r \over (q / \mu)^{1 / \nu_\parallel}}, v^*,
             {q \over \mu (g \omega / c \mu)^{1 / z}}, 1,
 {\omega^2 \over c^2 \mu^2 (g \omega / c \mu)^{2 (1 - \Delta)}} \right) \quad ,
\label{4.26}
\end{equation}
where \cite{Exp}
\begin{equation}
    2 - \eta_\parallel = (2 - \eta_\perp) / z  \, , \quad {\rm and} \quad
          \nu_\parallel = z \nu_\perp               \quad .
\label{4.28}
\end{equation}
Moreover, $\ell = r^{-1 / \zeta_r^*}$ leads to
\begin{equation}
     \Gamma_{11}(\mu,r,c,g,u,{\bf q},\omega) \propto r^\gamma
      {\hat \Gamma}_{11} \left( 1, v^*, {q \over \mu} r^{- \nu_\perp},
                 {g \omega \over c \mu} r^{- \nu_\parallel},
 {\omega^2 \over c^2 \mu^2} r^{-2 \nu_\parallel (1 - \Delta)} \right) \quad ,
\label{4.30}
\end{equation}
with another (``static'') exponent $\gamma$, related to $\nu_\perp$ and
$\nu_\parallel$ via
\begin{equation}
    \gamma = \nu_\perp (2 - \eta_\perp)
             = \nu_\parallel (2 - \eta_\parallel) \quad .
\label{4.31}
\end{equation}
For the sake of completeness, we finally state the scaling relations
\cite{Kin83}
\begin{equation}
     2 \beta + \gamma = 2 - \alpha = D \nu_\perp + \nu_\parallel
                       = 2 \beta \delta - \gamma = \beta (1 + \delta)
\label{4.32}
\end{equation}
for exponents $\alpha$, $\beta$, and $\delta$, to be defined in just the same
way as for ordinary continuous phase transitions \cite{Ami84}.

In the special case of isotropic percolation ($g_0 = 0$), the scaling relations
become considerably simpler. For example, instead of Eq.~(\ref{4.30}) one finds
{\it self--similar} behavior according to
\begin{equation}
      \Gamma_{11}(\mu,r,c,0,u,{\bf q},\omega) \propto r^\gamma
      {\hat \Gamma}_{11} \left( 1, v^*, {q \over \mu} r^{- \nu},0,
                      {\omega^2 \over c^2 \mu^2} r^{-2 \nu} \right) \quad ,
\label{4.33}
\end{equation}
with {\it two} independent critical exponents $\eta = - \zeta_\phi^*$ and $\nu
= -1 / \zeta_r^*$. The scaling relations reduce to $\gamma = \nu (2 - \eta)$
and $2 \beta + \gamma = 2 - \alpha = d \nu$. On leaving the self--similar
scaling region, a second scaling variable comes into play, leading to the
appearance of the crossover exponent $\Delta$ defined above [Eqs.~(\ref{4.20}),
(\ref{4.26}), and (\ref{4.30})], and eventually to anisotropic scaling.

%%%%%%%%%%%%%%%%%%%%%%%%%%%%%%%%%%%%%%%%%%%%%%%%%%%%%%%%%%%%%%%%%%%%%%%%%%%%%%%

\subsection{One--loop Results}

To one--loop order, Wilson's functions as derived from Eqs.~(\ref{4.2}) --
(\ref{4.6}) and (\ref{3.8}) -- (\ref{3.12}) read
\begin{eqnarray}
      \zeta_\phi &&= {v \over 8}
      \left[ 1 - {I^d_{35}(g) \over I^d_{17}(g)} \right] \quad , \label{4.34}\\
      \zeta_r    &&= -2 + {3 v \over 8}
           + {v \over 8} {I^d_{35}(g) \over I^d_{17}(g)} \quad , \label{4.35}\\
      \zeta_c    &&= - {v g^2 (d-8) \over 16}
         \left[ 2 {I^d_{37}(g) \over I^d_{17}(g)} -
                 {I^d_{55}(g) \over I^d_{17}(g)} \right] \quad , \label{4.36}\\
      \zeta_g    &&= -1 + {v \over 8}
         \left[ 1 - {I^d_{35}(g) \over I^d_{17}(g)} \right]
             - {v g^2 (d-8) \over 16}
         \left[ 2 {I^d_{37}(g) \over I^d_{17}(g)} -
                 {I^d_{55}(g) \over I^d_{17}(g)} \right] \quad , \label{4.37}\\
      \zeta_u  &&= {d-6 \over 2} + {13 v \over 16} + {3 v \over 16}
                   {I^d_{35}(g) \over I^d_{17}(g)}       \quad , \label{4.38}
\end{eqnarray}
[here Eqs.~(\ref{B5}) and (\ref{B6}) of Appendix B have been applied].

In the following two special situations, we may explicitly evaluate these
general relations: In the case $g \rightarrow 0$, we find $v \rightarrow 2 u^2
c$ [using (\ref{B3})]; thus
\begin{equation}
     g \rightarrow 0 \; : \qquad \beta_v =
                          v \left( d - 6 + {7 \over 4} v \right) \quad ,
\label{4.39}
\end{equation}
with the stable fixed point of isotropic percolation (for $d < 6$)
\begin{equation}
     v^*_{\rm I} = {4 \over 7} (6 - d) \quad .
\label{4.40}
\end{equation}
The fixed--point values for the $\zeta$ functions are $\zeta_\phi^* = v^*/12$,
$\zeta_r^* = -2 + 5 v^*/12$, $\zeta_c^* = 0$, and $\zeta_g^* = -1 + v^*/12$.
Hence we find the following isotropic critical exponents [$\eta =
\eta_\perp(v^*_{\rm I})$ and $\nu = \nu_\perp(v^*_{\rm I})$]
\begin{equation}
     \eta     = - {6 - d \over 21}   \, \quad {\rm and} \quad
      \nu^{-1} = 2 - {5 (6 - d) \over 21}         \quad .
\label{4.41}
\end{equation}
In Eq.~(\ref{4.20}), the term linear in $\omega$ vanishes for $g \rightarrow
0$, and inserting $\zeta_c^* = 0$ yields
\begin{equation}
     z (1 - \Delta) = 1 \quad ,
\label{4.43}
\end{equation}
which is equivalent to isotropic scaling in $d = D + 1$ dimensions; compare
(\ref{4.33}). The crossover exponent
\begin{equation}
     \Delta = {1 - (6 - d) / 21 \over 2 - (6 - d) / 21}
\label{4.44}
\end{equation}
provides the power law according to which the isotropic scaling region is left
in favor of the anisotropic behavior to be discussed below. For $d > 6$, the
isotropic Gaussian fixed point
\begin{equation}
     v^*_{\rm GI} = 0
\label{4.29}
\end{equation}
is stable, with the corresponding mean--field exponents
\begin{equation}
     \eta_\perp = 0  \, , \quad \nu_\perp = {1 \over 2}  \, , \quad
                z = 2 \, , \quad {\rm and} \quad \Delta = {1 \over 2} \, .
\label{4.47}
\end{equation}

The limit of directed percolation is characterized by a diverging anisotropy
scale $g \rightarrow \infty$. Using Eq.~\ref{B4}, the coupling parameter
becomes $v \rightarrow u^2 c B(1/2,(7-d)/2)/g$ (note the additional factor $1 /
g$), and the $\beta$ function now reads
\begin{equation}
   g  \rightarrow \infty \; : \qquad \beta_v =
                          v \left( d - 5 + {3 \over 2} v \right) \quad .
\label{4.45}
\end{equation}
Therefore, for $d > 5$ the directed Gaussian fixed point
\begin{equation}
     v^*_{\rm GD} = 0
\label{4.46}
\end{equation}
becomes stable, which again leads to the exponents of Eq.~(\ref{4.47}).

On the other hand, for dimensions $d < 5$ the asymptotic behavior is governed
by the non--trivial anisotropic scaling fixed point
\begin{equation}
     v^*_{\rm D} = {2 \over 3} (5 - d) \quad .
\label{4.51}
\end{equation}
Note that the upper critical dimension has changed to $d_c^{\rm D} = 5$, in
contrast to the isotropic case, where $d_c^{\rm I} = 6$; physically, this
reduction is due to the suppression of fluctuations by the anisotropy, and may
be formally traced back to the appearance of the factor $1/g$ in the expression
for the asymptotic coupling. Inserting (\ref{4.51}) into the results for the
$\zeta$ functions, $\zeta_\phi^* = v^* / 8$, $\zeta_r^* = -2 + 3 v^* / 8$,
$\zeta_c^* = v^* / 8$, and $\zeta_g^* = -1 + v^* / 4$, we find the following
critical indices
\begin{equation}
     \eta_\perp    = - {5 - d \over 12}  \, , \quad
     \nu_\perp^{-1} = 2 - {5 - d \over 4}   \, , \quad
     z = 2 - {5 - d \over 12}  \,  ,  \quad  {\rm and} \quad
     z \Delta = 1 - {5 - d \over 6} \quad ,
\label{4.52}
\end{equation}
which characterize the self--affine scaling of the percolation clusters with
preferred ``time'' direction \cite{Car80}. The anisotropy is reflected in the
exponent $z$, and $\Delta$ now describes the crossover from isotropic to
directed percolation near the anisotropic percolation fixed point $v^*_{\rm
D}$. We have collected the four fixed points and the corresponding values for
the independent critical exponents in Table I.

Thus we have demonstrated that both the self--similar and the self--affine
scaling behavior are within the scope of the present theory, at least for
dimensions $d \leq 5$; for $5 < d \leq 6$ the model is not renormalizable in
the directed limit, and the crossover description based on extracting the UV
poles may be questionable. However, in this case the asymptotics of the model
are simply described by the mean--field exponents corresponding to the Gaussian
fixed point $v^*_{\rm GD}$, with logarithmic corrections for $d = 5$, and at
least the qualitative features of the crossover to this Gaussian theory are
well reproduced by our formalism (see also Refs. \cite{Fre88} and
\cite{Tau93}). At any rate, this discussion again emphasizes the fact that no
expansion with respect to a {\it fixed} upper critical dimension can be applied
consistently (from our procedure a kind of ``effective upper critical
dimension'' $d_c(g)$, varying with the anisotropy scale $g$, might be
extracted).

For the physically interesting situation $d < 5$, note that a smooth
interpolation between the two asymptotic cases is obtained at every stage of
the present theory, i.e., for the renormalization constants, for the $\zeta$
and $\beta$ functions, for the fixed--point values and the scaling functions.
We shall now proceed to study the entire crossover region between these
asymptotic regimes, which can be readily done by solving the coupled set of
flow equations (\ref{4.8}) with (\ref{4.34}) -- (\ref{4.38}) numerically.

We start with the analysis of the flow diagram for the effective coupling
``constants'' $v(\ell)$ and ${\tilde g}(\ell) = g(\ell) / [ 1 + g(\ell)]$ (the
latter assumes values in the interval $[0,1]$ only, which is more convenient
than the range $[0,\infty]$ of the original anisotropy parameter $g$),
shown in Fig. 5. The four fixed points, as summarized in Table I, namely those
for isotropic percolation (I), directed percolation (D), Gaussian isotropic
(GI), and Gaussian directed (GD) percolation, respectively, determine the
topology of the $(v,{\tilde g})$ flow diagram (where we have chosen $d = 3$ and
$\mu = 1$ for the renormalization scale.) The only infrared--stable fixed point
is the one for directed percolation, $(v^*,{\tilde g}^*)=({2 \over 3}
(5-d),1)$. All other fixed points are unstable, but as can be inferred from
Fig. 5, they are more or less attractive depending on the initial value for the
coupling constants. The flow diagram is divided into two regions by a
separatrix, which constitutes the renormalization--group trajectory from the
fixed point $v^*_{\rm I}$ to $v^*_{\rm D}$, describing the universal crossover
from self--similar to self--affine scaling. For initial values $v \leq v_{\rm
I}$, there are besides the stable fixed point for directed percolation three
unstable fixed points (I), (GI), and (GD). Starting from the Gaussian isotropic
fixed point (GI), the renormalization--group trajectories traverse regions
close to the fixed points (I) or (GD), depending on the initial values ${\tilde
g(1)}$ and $v(1)$, before they finally reach the infrared--stable fixed point
for directed percolation (D). The competition between all these fixed points
will also become apparent in the flow of the effective exponents (e.g., for the
pair correlation function). For initial values $v\geq v^*_{\rm I}$ there is
only one relevant unstable fixed point (I).

As a result of the flow dependence of the coupling constants also the $\zeta$
functions, whose fixed--point values are related to the critical exponents,
display crossover behavior. In Fig. 6 we exemplify this flow dependence for the
$\zeta$ functions $\zeta_\phi(\ell)$ and $\zeta_r(\ell)$, respectively, where
we have chosen a fixed initial value $v(1) = v^*_{\rm I}$ for the three--point
coupling, and a series of initial values for the anisotropy parameter [$g(1) =
10^{-2}$, $10^{-3}$, and $10^{-4}$].

The most important conclusion to be drawn from the crossover behavior of the
$\zeta$ functions is that the crossover for the anomalous dimension of the
stochastic fields $\zeta_\phi$ starts at values of the flow parameter which are
approximately one order of magnitude smaller than the corresponding values for
$\zeta_r$. This should then also be visible as different crossover locations
for the distinct effective exponents of the connectivity $G$.

%%%%%%%%%%%%%%%%%%%%%%%%%%%%%%%%%%%%%%%%%%%%%%%%%%%%%%%%%%%%%%%%%%%%%%%%%%%%%%%

\subsection{Effective Exponents for the Pair Correlation Function}

In this paragraph we consider the effects of the crossover on the most
interesting physical quantity, namely the pair correlation function. As
discussed in Sec. II, the scaling behavior of the connectivity is identical to
that of the two-point function $G_{11}({\bf q}, \omega) = \Gamma_{11}(-{\bf q},
-\omega)^{-1}$, which has been studied in the previous subsection.

As a first approximation, one can use the zero--loop result for the scaling
function and finds
\begin{equation}
     \Gamma_{11}(r,{\bf q},\omega) =
      \mu^2 \ell^2 e^{\int_1^\ell \zeta_{\phi}(\ell') d\ell' / \ell'}
      \left[ r(\ell) + {q^2 \over \mu^2 \ell^2} +
             {\omega^2 \over \mu^2 \ell^2 c(\ell)^2} +
             2 i {\omega g (\ell) \over \mu \ell c(\ell)} \right] \quad .
\label{4.53}
\end{equation}
It is quite straightforward to calculate corrections to this zero--loop result
for the scaling function in a perturbation expansion with respect to the
effective coupling constant $v$. However, there are serious technical problems
associated with the non--analytical dependence of the shift of the percolation
threshold $p_c$ on $u_0$ [Eq.~(\ref{3.15})]. This is, however, not a particular
difficulty of our theoretical approach, but a fundamental problem for any
field--theoretical calculation at fixed dimension below $d_c$. Schloms and Dohm
\cite{Doh89} have shown that for a $\phi^4$ theory the nonvanishing mass shift
can be incorporated in the minimal subtraction approach directly in three
dimensions without recourse to the standard ($\epsilon = 4-d$) expansion. But,
even for this ``standard model'' of critical phenomena there does not exist a
straightforward perturbation method at finite external momenta which allows for
a consistent treatment of the non--analytical mass shift at fixed dimension.
Hence, at the present stage of the theory, we can treat the $\zeta$ functions
to arbitrary loop--order, but have to restrict ourselves to a mean--field
(zero--loop) treatment for the scaling functions. Nevertheless, one expects
that the results for the effective exponents obtained within this
``renormalized mean--field theory'' to provide a reasonably good approximation.
This expectation is based on the experience that amplitude (or scaling)
functions are usually smooth and well--behaved, and enter the results less
sensitively than the exponential $\zeta$ functions. In fact, our approach was
especially designed to incorporate the complete crossover behavior into the
exponential functions $\zeta_a$. This has to be contrasted with the usual
scheme, where the crossover has to be inferred from high--order calculations of
the amplitude functions (and usually does not go far beyond the determination
of corrections to scaling).

The most convenient way to analyze the crossover behavior of the pair
correlation function is in terms of effective critical exponents. We consider
first the case $r=0$ and ${\bf q} = 0$ and choose to define the effective
critical exponent $\eta_{\parallel \, {\rm eff}}$ by [compare (\ref{4.26})]
\begin{equation}
     2 - \eta_{\parallel \, \rm eff}(\omega) =
      {d \ln {\sqrt{\mid \Gamma_{11}(0,{\bf 0},\omega) \mid^2 }}
       \over d \ln \omega}       \quad .
\label{4.54}
\end{equation}
Using the matching condition
\begin{equation}
     \left \vert {\omega^2 \over \mu^2 \ell^2 c^2 (\ell)} +
      2 i {\omega g (\ell) \over \mu \ell c(\ell)} \right\vert^2 = 1 \quad ,
\label{4.55}
\end{equation}
the effective exponent is found to be
\begin{equation}
    2 - \eta_{\parallel \, \rm eff}(\ell) =
      \left[ 2 + \zeta_{\phi}(\ell) \right] \,
        {d \ln \ell \over d \ln \omega }  \quad ,
\label{4.56}
\end{equation}
where the factor $d \ln \ell / d \ln \omega$ has to be determined from
\ref{4.55}. The effective exponent $2 - \eta_{\parallel \, \rm eff}(\ell)$ is
shown in Fig. 7 as a function of the flow parameter $\ell$, where for the
initial value $v(1) = v^*_{\rm I}$, and a series of initial values for the
anisotropy parameter, $g(1) = 10^{-k}$ with $k=2,3,4$, have been chosen. Upon
using the matching condition (\ref{4.55}) the flow parameter $\ell$ can be
related to the longitudinal length scale (``frequency''). If the effective
exponent is plotted versus the running anisotropy parameter $g(\ell)$, all the
curves for different initial values $g(1)$ collapse onto one master curve. The
corresponding plot is shown in Fig. 8. Since $g(\ell) \propto
\ell^{\zeta_g^*}$ near a fixed point, the scale transformation (in order to
reduce all curves to one master curve) depends on how close the flow is to one
of the four different fixed points. Note that because $\zeta_g$ is not a
constant, no simple scaling relation can be deduced for the location of the
crossover by simply investigating any one of the two asymptotic regimes.

Next, in the case $r=0$ and $\omega = 0$ one can define an effective
exponent $\eta_{\perp \, {\rm eff}}$ by [see Eq.~(\ref{4.20})]
\begin{equation}
      2 - \eta_{\perp \, {\rm eff}}(q) = { d \ln \Gamma_{11}(0,{\bf q},0)
      \over d \ln q} \quad .
\label{4.59}
\end{equation}
With the matching condition $(q/\mu \ell)^2 = 1$ this reduces to
\begin{equation}
    2- \eta_{\perp \, {\rm eff}}(\ell) = 2 + \zeta_\phi (\ell) \quad .
\label{4.60}
\end{equation}
Finally, we consider the case $\omega = 0$ and ${\bf q} = 0$. Upon defining an
effective exponent [compare (\ref{4.30})]
\begin{equation}
    \gamma_{\rm eff}(r) = {d \ln \Gamma_{11}(r,{\bf 0},0) \over d \ln r}
\label{4.57}
\end{equation}
and choosing the matching condition $r(\ell) = 1$ we find
\begin{equation}
     \gamma_{\rm eff}(\ell) = \left[ 2 + \zeta_\phi (\ell) \right]
                               { d \ln \ell \over d \ln r}
      = - { 2 + \zeta_\phi (\ell) \over \zeta_r (\ell) } \quad ;
\label{4.58}
\end{equation}
remarkably, Eq.~(\ref{4.58}) is valid throughout the entire crossover region,
and not just near one of the fixed points, where it becomes identical to the
scaling relation (\ref{4.31}).

The flows of the above effective exponents, and that of the ``effective''
dynamic exponent
\begin{equation}
     z_{\rm eff} (\ell) = 1 + \zeta_c (\ell) - \zeta_g (\ell) \quad ,
\label{4.61}
\end{equation}
are shown in Figs. 8 -- 11 versus the scaling variable $\ln [g(\ell)]$, with
$\mu = 1$, and the initial value $v(1) = v^*_{\rm I}$, such that the universal
crossover starting from the isotropic scaling region is described. The most
important conclusion to be drawn from these results is that the characteristic
anisotropy scales $g(\ell_{\rm cross})$, where the crossover occurs for the
effective exponents defined above, are different ! The effective exponent
$\eta_{\parallel \, \rm eff}$ starts to cross over from the isotropic to the
directed fixed point value already at $\ln g(\ell_{\rm cross}) \approx -0.8$,
whereas the effective exponent $\gamma_{\rm eff}$ shows this crossover at
$\ln g(\ell_{\rm cross}) \approx -0.1$, and the ``dynamical'' exponent $z_{\rm
eff}$, as well as $\eta_{\perp \, \rm eff}$, at an even larger value $\ln
g(\ell_{\rm cross}) \approx 0.8$. Note that the remarkable change of
$\eta_{\parallel \, \rm eff}$ is already apparent at mean--field level, where
it acquires the values $0$ and $1$ in the isotropic and directed limit,
respectively. However, a description of the crossovers for $\eta_{\perp \, \rm
eff}$, $\gamma_{\rm eff}$, and $z_{\rm eff}$ requires the $\zeta$ functions as
least on the one--loop level, as has been achieved here.

We remark that a precise calculation of these crossover features has not been
possible up to this present work. Of course, the exponents for the
limits of both isotropic and directed percolation have been determined to a
much higher accuracy than is provided in our one--loop approximation
\cite{Ess80,Kin83,Car80,Bro78,Hen90,Gra89,Car84,Gra79,Ben84}. In fact, the
relative errors of our one--loop results, as compared to the values given in
Table III of Ref. \cite{Kin83}, are approximately $0.03$, $0.09$, and $0.21$
for the exponent $\gamma$ of directed percolation in $d=4$, $3$, and $2$
dimensions, respectively, and for $\nu_\perp$ one finds correspondingly $0.10$
and $0.27$ for the three-- and two--dimensional cases. In the isotropic limit
at two dimensions, the relative errors are $0.21$ and $0.27$ for $\gamma$ and
$\nu$, respectively. (We remark that the numerical values of our one--loop
results are even slightly better than those of an additional $\epsilon$
expansion; yet they also improve as the upper critical dimension is
approached.) However, our aim was rather to calculate the detailed crossover
properties, and we believe that the characteristic crossover scales
$g(\ell_{\rm cross})$ should not be affected too severely by, e.g., higher
orders of perturbation theory. Certainly, these results, i.e., the
approximations used in this paper, are subject to tests by both computer
simulations and experiment. Of course, it would be very interesting to compare
our predictions concerning the crossover scales of the different effective
exponents with the outcome of such numerical simulations and/or physical
experimental setups.

Therefore we add some remarks on the interpretation of our results, and on the
number of free parameters of the theory. One of the advantages of studying the
effective exponents, is the fact that they do {\it not} depend on
non--universal amplitudes, i.e., on the initial values $u(1)$, $c(1)$ (which
can be set to $1$, if one starts from the isotropic scaling region), and the
scale $\mu$. For convenience, we have plotted our results versus the scaling
variable $\ln [g(\ell)]$ in Figs. 8 -- 11, and have thus also eliminated the
dependence on the formal anisotropy parameter. However, in simulations or
experiments $g(1) = g$ is a fixed quantity, although its correspondence to a
physical anisotropy measure may be rather indirect. In any case, if merely the
{\it universal} crossover features are to be investigated, the initial value of
the effective coupling should be chosen as $v(1) = v^*_{\rm I}$, in order to
resemble the self--similar scaling behavior, and then $g(1)$ is the {\it only}
free parameter of the theory. In order to compare directly with our figures,
one then has to solve the flow equations (\ref{4.8}), (\ref{4.15}) with
Eqs.~(\ref{4.34}) -- (\ref{4.38}), and apply the relevant matching condition,
which is straightforward (unfortunately, the scaling behavior with respect to
$g(1)$ is complex and cannot be described by a pure power law). But once $g(1)$
has been determined for one of the effective exponents, it must necessarily
also yield the crossover point for any of the others. Obviously, a quantitative
study of the effective indices $\eta_{\parallel \, \rm eff}(\omega)$ and
$\gamma_{\rm eff}(r)$ is most promising, while a detailed analysis of
$\eta_{\perp \, \rm eff}(q)$ and $z_{\rm eff}(\ell)$ is probably difficult,
because of their comparatively small changes on approaching the self--affine
regime.

Possibly clearer on first sight, however somewhat less distinct concerning
quantitative features, is a discussion of the crossover from self--similar to
self--affine scaling on the basis of contour plots, e.g., for
$\Gamma_{11}(0,{\bf q},\omega)$. Applying the matching condition
\begin{equation}
     \left \vert {q^2 \over \mu^2 \ell^2} +
      {\omega^2 \over \mu^2 \ell^2 c(\ell)^2} +
      2 i {\omega g(\ell) \over \mu \ell c(\ell)} \right \vert = 1 \quad ,
\label{4.62}
\end{equation}
valid at the percolation threshold $p = p_c$, Eq.~(\ref{4.53}) simply becomes
\begin{equation}
     \Gamma_{11}(0,{\bf q},\omega) = \mu^2 \ell^2
               e^{\int_1^\ell \zeta_\phi(\ell') d\ell' / \ell'} \quad ,
\label{4.63}
\end{equation}
which is a monotonic function of $\ell$ (the exponential is approximately given
by $\ell^{-\eta_{\perp \, \rm eff}(\ell)}$, see Fig. 9). Therefore, contours of
constant $\Gamma_{11}$ are identical to contours of constant flow parameter
$\ell$, which in turn can be easily inferred from Eq.~(\ref{4.62}). In Fig. 12,
we depict typical examples of such contour plots, with values of $\ell =
10^{-4}$, $10^{-3}$, and $10^{-2}$, respectively, from which the crossover from
isotropic to anisotropic scaling becomes apparent.

%%%%%%%%%%%%%%%%%%%%%%%%%%%%%%%%%%%%%%%%%%%%%%%%%%%%%%%%%%%%%%%%%%%%%%%%%%%%%%%

\section{Summary and Conclusion}

In this paper we have investigated the crossover from isotropic to directed
percolation taking advantage of a mapping onto a field--theoretical
representation of the connectivity \cite{Car80,Car84}. From a conceptual point
of view, the main result of the present paper is the demonstration that an
extended minimal subtraction scheme is capable of dealing with crossover
problems associated with a change in the upper critical dimension $d_c$ in the
framework of renormalization group theory. We have exemplified the method for a
long standing problem in percolation theory. But we believe that this approach
can be applied to a wide variety of interesting physical problems. Among those
of most current interest are the crossover between bulk and surface physics
\cite{Die86}, between mean--field and critical behavior \cite{Tau93}, crossover
from propagating to overdamped soft modes in critical dynamics, e.g., near
structural phase transitions, and others.

Since our primary goal was to present the formalism, we have restricted
ourselves to a one--loop approximation. We have identified the crossover
exponent $\Delta$ and calculated effective exponents for different length
scales and the pair correlation function. Higher loop orders for the
exponential ($\zeta$) functions are in principle accessible within our
theoretical framework. However, the calculation of the wavevector and
frequency dependence of the amplitude functions and the incorporation of the
non--analytical mass shift remains an open problem for a field theoretical
calculation at fixed dimension \cite{Doh89}.

It would also be interesting to investigate the crossover from isotropic to
directed percolation by numerical simulations or physical experiments, and
compare them with the theoretical results of our paper. Even though the values
of the critical exponents in the two asymptotic regimes are not accurate
(one--loop results), we expect the crossover behavior of the effective
exponents to be qualitatively correct. Especially the predictions concerning
the different loci of the crossover for the ``static'' and ``dynamic''
quantities could be tested by numerical simulations. Obviously, this would be
of considerable help for estimating the quality of the proposed approach to
crossover phenomena also in different situations, where a numerical simulation
is either very cumbersome or not feasible at all.

\acknowledgments

We would like to thank S. Clar, who simulated the percolation clusters of
Fig. 1. We gratefully acknowledge support from the Deutsche
Forschungsgemeinschaft (DFG) under Contracts No. Fr. 850/2-1, Ta. 177/1-1,
and Schw. 348/4-2.
%\newpage

%%%%%%%%%%%%%%%%%%%%%%%%%%%%%%%%%%%%%%%%%%%%%%%%%%%%%%%%%%%%%%%%%%%%%%%%%%%%%%%

\section{Appendix}

\subsection{Two-- and Three--point Vertex Functions to One--loop Order}

For the sake of completeness, we list some important explicit analytical
results for the two-- and three--point vertex functions to one-loop order. The
corresponding Feynman diagrams are depicted in Fig. 4. Writing the free
propagator in the form
\begin{equation}
  G_{11 \, (0)}^0(-{\bf q},-\omega) = {c_0^2 \over [\omega - \omega_+({\bf q})]
                                         [\omega - \omega_-({\bf q})]} \quad ,
\label{A1}
\end{equation}
where $\omega_\pm({\bf q})$ represents the ``dispersion relation'' for the
``elementary excitations'' of our model,
\begin{equation}
  \omega_\pm({\bf q}) = -i c_0 \left( g_0 \mp \sqrt{r_0 + g_0^2 +q^2} \right)
                                                                    \quad ,
\label{A2}
\end{equation}
one finds for $\Gamma_{11}^0({\bf q},\omega)$:

\begin{eqnarray}
     (a) &&= r_0 + q^2 + \omega^2 / c_0^2 + 2 g_0 i \omega / c_0
                                                          \quad , \label{A3}\\
      (b) &&= {u_0^2 c_0^4 \over 2} \int_k \int_\nu
               G_{11 \, (0)}^0({\bf k} - {\bf q} / 2,\nu - \omega / 2)
        G_{11 \, (0)}^0({\bf k} + {\bf q} / 2,\nu + \omega / 2)    \label{A4}\\
          &&= u_0^2 c_0^5 \int_k {\scriptstyle
              {1 \over \sqrt{r_0 + g_0^2 + ({\bf q} / 2 - {\bf k})^2}}
 {({\bf q} {\bf k}) + \omega^2 / 2 c_0^2 + 2 g_0 i \omega / c_0 - 2 g_0^2 \over
 \left[ -\omega + \omega_+({\bf q} / 2 + {\bf k}) +
                                        \omega_+({\bf q} / 2 - {\bf k}) \right]
   \left[ -\omega + \omega_-({\bf q} / 2 + {\bf k}) +
                \omega_+({\bf q} / 2 - {\bf k}) \right]}}   \times \label{A5}\\
&&\qquad \qquad \qquad \qquad \qquad \times {\scriptstyle {1 \over
   \left[ -\omega + \omega_+({\bf q} / 2 + {\bf k}) +
                                        \omega_-({\bf q} / 2 - {\bf k}) \right]
   \left[ -\omega + \omega_-({\bf q} / 2 + {\bf k}) +
                \omega_-({\bf q} / 2 - {\bf k}) \right]}}    \quad . \nonumber
\end{eqnarray}
In the last step, we have performed the integration over the internal frequency
$\nu$ via the residue theorem.

As a special case, we list the result for vanishing external wavevector,
${\bf q} = 0$, which enters the calculation of the effective exponents:
\begin{equation}
     \Gamma_{11}^0({\bf 0},\omega)
      = r_0 + {\omega^2 \over c_0^2} + 2 g_0 {i \omega \over c_0}
      + {u_0^2 c_0 \over 8} \int_k {1 \over \sqrt{r_0 + g_0^2 +k^2}}
      {1 \over r_0 + k^2 + {\omega^2 \over 4 c_0^2} + g_0 {i \omega \over c_0}}
                                                                      \quad .
\label{A6}
\end{equation}

In order to calculate the renormalization constants, we may expand (\ref{A5})
with respect to $q^2$ and $\omega$:
\begin{eqnarray}
     \Gamma_{11}^0({\bf q},\omega) &&= r_0 \Biggl( 1 + {u_0^2 c_0 \over 8 r_0}
     \int_k {1 \over \sqrt{r_0 + g_0^2 + k^2} (r_0 + k^2)} \Biggr) \nonumber\\
      &&+ 2 g_0 {i \omega \over c_0} \Biggl( 1 - {u_0^2 c_0 \over 16}
   \int_k {1 \over \sqrt{r_0 + g_0^2 + k^2} (r_0 + k^2)^2} \Biggr) \label{A7}\\
      &&+ {\omega^2 \over c_0^2} \Biggl( 1 - {u_0^2 c_0 \over 32}
    \int_k {1 \over \sqrt{r_0 + g_0^2 + k^2} (r_0 + k^2)^2}
                                         - {u_0^2 c_0 g_0^2 \over 8}
    \int_k {1 \over \sqrt{r_0 + g_0^2 + k^2} (r_0 + k^2)^3} \Biggr) \nonumber\\
      &&+ q^2 \Biggl( 1 - {u_0^2 c_0 \over 32}
 \int_k {1 \over \sqrt{r_0 + g_0^2 + k^2} (r_0 + k^2)^2} - {u_0^2 c_0 \over 64}
  \int_k {1 \over \sqrt{(r_0 + g_0^2 + k^2)^3} (r_0 + k^2)} \Biggr) \nonumber\\
      &&+ {u_0^2 c_0 \over 32} \int_k {({\bf q} {\bf k})^2
                             \over \sqrt{(r_0 + g_0^2 + k^2)^3} (r_0 + k^2)^2}
        + {3 u_0^2 c_0 \over 64} \int_k {({\bf q} {\bf k})^2
                    \over \sqrt{(r_0 + g_0^2 + k^2)^5} (r_0 + k^2)} \nonumber\\
                                  &&+ {\cal O}(\omega^3,q^4) \quad. \nonumber
\end{eqnarray}

Similarly, the three--point vertex function
$\Gamma_{12}^0(-{\bf q},-\omega; {\bf q} / 2, \omega / 2; {\bf q} / 2,
\omega / 2)$ reads
\begin{eqnarray}
     (c) &&= - u_0 / 2                                   \quad , \label{A8}\\
      (d) &&= u_0^3 c_0^6 \int_k \int_\nu
             G_{11 \, (0)}^0({\bf k} - {\bf q} / 2,\nu - \omega / 2)
             G_{11 \, (0)}^0({\bf k} + {\bf q} / 2,\nu + \omega / 2)
             G_{11 \, (0)}^0 ({\bf k},\nu)                       \label{A9}\\
         &&= {u_0^3 c_0^5 \over 2} \int_k {\scriptstyle \Biggl[
             {1 \over \sqrt{r_0 + g_0^2 + ({\bf q} / 2 + {\bf k})^2}} {1 \over
   \left[ -\omega + \omega_+({\bf q} / 2 + {\bf k}) +
                                       \omega_+({\bf q} / 2 - {\bf k}) \right]
   \left[ -\omega + \omega_+({\bf q} / 2 + {\bf k}) +
        \omega_-({\bf q} / 2 - {\bf k}) \right]}}        \times \nonumber\\
&&\qquad \qquad \qquad \qquad \qquad \times {\scriptstyle {1 \over
   \left[ -\omega / 2 + \omega_+({\bf q} / 2 + {\bf k}) -
                                                     \omega_+({\bf k}) \right]
   \left[ -\omega / 2 + \omega_+({\bf q} / 2 + {\bf k}) -
        \omega_-({\bf k}) \right]}}                             \nonumber\\
&&\qquad \qquad + {\scriptstyle
             {1 \over \sqrt{r_0 + g_0^2 + ({\bf q} / 2 - {\bf k})^2}} {1 \over
   \left[ -\omega + \omega_+({\bf q} / 2 + {\bf k}) +
                                       \omega_-({\bf q} / 2 - {\bf k}) \right]
   \left[ -\omega + \omega_-({\bf q} / 2 + {\bf k}) +
        \omega_-({\bf q} / 2 - {\bf k}) \right]}}        \times \nonumber\\
&&\qquad \qquad \qquad \qquad \qquad \times {\scriptstyle {1 \over
   \left[ -\omega / 2 + \omega_-({\bf q} / 2 - {\bf k}) +
                                                     \omega_+({\bf k}) \right]
   \left[ -\omega / 2 + \omega_-({\bf q} / 2 - {\bf k}) +
        \omega_-({\bf k}) \right]}}                            \label{A10}\\
&&\qquad \qquad + {\scriptstyle {1 \over \sqrt{r_0 + g_0^2 + k^2}} {1 \over
   \left[ -\omega / 2 + \omega_+({\bf q} / 2 + {\bf k}) -
                                                     \omega_+({\bf k}) \right]
   \left[ -\omega / 2 + \omega_-({\bf q} / 2 + {\bf k}) -
        \omega_+({\bf k}) \right]}}                       \times \nonumber\\
&&\qquad \qquad \qquad \qquad \qquad \times {\scriptstyle {1 \over
   \left[ -\omega / 2 + \omega_+({\bf q} / 2 - {\bf k}) +
                                                     \omega_-({\bf k}) \right]
   \left[ -\omega / 2 + \omega_-({\bf q} / 2 - {\bf k}) +
        \omega_-({\bf k}) \right]}}                \Biggr] \quad . \nonumber
\end{eqnarray}

For the determination of the renormalization constant $Z_u$, however, we merely
need
\begin{eqnarray}
     &&\Gamma_{12}^0({\bf 0},0;{\bf 0},0;{\bf 0},0) =
      - \Gamma_{21}^0({\bf 0},0;{\bf 0},0;{\bf 0},0)             \label{A11}\\
      &&= - {u_0 \over 2} \Biggl( 1 -
 {u_0^2 c_0 \over 4} \int_k {1 \over \sqrt{r_0 + g_0^2 +k^2} (r_0 + k^2)^2} -
 {u_0^2 c_0 \over 8} \int_k {1 \over \sqrt{(r_0 + g_0^2 + k^2)^3} (r_0 + k^2)}
                                                    \Biggr) \quad . \nonumber
\end{eqnarray}
%

%\newpage

%%%%%%%%%%%%%%%%%%%%%%%%%%%%%%%%%%%%%%%%%%%%%%%%%%%%%%%%%%%%%%%%%%%%%%%%%%%%%%%

\subsection{Feynman Parameter Integrals}

Within the generalized minimal subtraction scheme described above, the
determination of the Z factors after performing the frequency and momentum
integration using the Feynman parametrization
\begin{equation}
     {1 \over A^r B^s} = {\Gamma(r+s) \over \Gamma(r) \Gamma(s)} \int_0^1
      {x^{r-1} (1-x)^{s-1} \over [x A + (1-x) B]^{r+s}} dx
\label{B1}
\end{equation}
[$\Gamma(x)$ is Euler's Gamma function] leads to integrals of the form (with
odd integers $m,n$)
\begin{equation}
     I^d_{mn}(g) = \int_0^1 {x^{m/2 - 1} \over (1 + x g^2)^{(m+n-d)/2}} dx
           = g^{-m} \int_0^{g^2} {t^{m/2 - 1} \over (1 + t)^{(m+n-d)/2}} dt
      \quad ,
\label{B2}
\end{equation}
for which one may immediately derive asymptotic relations for $g = 0$ and $g
\rightarrow \infty$, respectively [$B(x,y) = \Gamma(x) \Gamma(y) /
\Gamma(x+y)$ denotes Euler's Beta function],
\begin{eqnarray}
      I^d_{mn}(0) &&= 2 / m                                \quad , \label{B3}\\
      \lim_{g \rightarrow \infty} \left[ g^m I^d_{mn}(g) \right] &&=
           B \left( {m \over 2},{n-d \over 2} \right) =
      {\Gamma({m \over 2}) \Gamma({n-d \over 2}) \over \Gamma({m+n-d \over 2})}
                                                           \quad . \label{B4}
\end{eqnarray}
For the calculation of the $\zeta$ functions, the following formulas are very
useful:
\begin{eqnarray}
     I^d_{mn}(g) - g^2 I^d_{m+2,n}(g) &&= I^d_{m,n+2}(g) \quad , \label{B5}\\
      \mu {\partial \over \partial \mu} I^d_{mn} (g_0 / \mu) &&=
      (m + n - d) {g_0^2 \over \mu^2} I^d_{m+2,n} (g_0 / \mu)
                                                          \quad . \label{B6}
\end{eqnarray}
%

%%%%%%%%%%%%%%%%%%%%%%%%%%%%%%%%%%%%%%%%%%%%%%%%%%%%%%%%%%%%%%%%%%%%%%%%%%%%%%%

%%%%%%%%%%%%%%%%%%%%%%%%%%%%%%%%%%%%%%%%%%%%%%%%%%%%%%%%%%%%%%%%%%%%%%%%%%%%%%%

% the tables follow here
\begin{table}
\setdec 0.00
\caption{Fixed point values for the effective coupling ``constants'' $v(\ell)$
and $\tilde g(\ell)$ at the isotropic (I), Gaussian isotropic (GI), directed
(D), and Gaussian directed (GD) fixed points. and the corresponding values of
the four independent critical exponents. The other critical indices may be
inferred from scaling relations.}
\smallskip
\mediumtext

\begin{tabular}{lcccccc}
{\rm Fixed point}    &  $v^*$ &  ${\tilde g}^*$  &  $\eta_\perp$ &
$\nu_\perp^{-1}$  &  $z$  & $\Delta$      \\
\tableline
{\rm GI}     &  $0$ & $0$ & $0$ & ${2}$ & $2$ & ${1 \over 2}$ \\
{\rm I}      & ${4 \over 7} (6-d)$ & $0$ & $-{6-d \over 21}$ & $2 - {5 (6-d)
\over 21}$ & $2 - {6-d \over 21}$ & ${1 - (6-d)/21 \over 2 - (6-d)/21}$  \\
{\rm GD}     &  $0$ & $1$ & $0$ & ${2}$ & $2$ & ${1 \over 2}$ \\
{\rm D}      &  ${2 \over 3} (5-d)$ & $1$ & $- {5-d \over 12}$ & $2 - {5-d
\over 4}$ & $2 - {5-d \over 12}$ & ${1-(5-d)/6 \over 2 - (5-d)/12}$ \\
\end{tabular}
\label{table1}
\end{table}

%%%%%%%%%%%%%%%%%%%%%%%%%%%%%%%%%%%%%%%%%%%%%%%%%%%%%%%%%%%%%%%%%%%%%%%%%%%%%%%
% the figures follow here

\figure{Fig.1: Isotropic (a) and directed (b) percolation clusters.
\label{fig1}}

\figure{Fig.2: Directed bonds placed on a square percolation lattice.
\label{fig2}}

\figure{Fig.3: Propagators and vertices of the field--theoretical
representation used in the text.
\label{fig3}}

\figure{Fig.4: One-loop diagrams for (a) $\Gamma_{11}^0$, (b) $\Gamma_{12}^0$,
and (c) $\Gamma_{21}^0$.
\label{fig4}}

\figure{Fig.5: Renormalization--group trajectories in the $(v,{\tilde g})$
parameter space for the crossover from isotropic to directed percolation (to
one--loop order) in $d = D + 1 = 3$ dimensions. In order to map the range
$[0,\infty]$ of $g$ onto the interval $[0,1]$, we use the variable ${\tilde g}
= g /(1 + g)$ instead of the anisotropy parameter $g$.
\label{fig5}}

\figure{Fig.6: Flow dependence of $\zeta_\phi(\ell)$ (a) and $\zeta_r(\ell)$
(b), where we have chosen a fixed initial value $v(1) = v^*_{\rm I}$ for the
three--point coupling and a series of initial values for the anisotropy
parameter [$g(1) = 10^{-2}$, $10^{-3}$, and $10^{-4}$].
\label{fig6}}

\figure{Fig.7: Effective exponent $2 - \eta_{\parallel \, {\rm eff}}(\ell)$ for
the connectivity as a function of the flow parameter $\ell$ for fixed initial
values $v(1) = v^*_{\rm I}$ and a series of initial values for the
anisotropy parameter [$g(1) = 10^{-2}$, $10^{-3}$, and $10^{-4}$].
\label{fig7}}

\figure{Fig.8: Effective exponent $\eta_{\parallel \, {\rm eff}}(\ell)$ for the
pair correlation function as a function of the running anisotropy parameter
$g(\ell)$ for a fixed initial value of the coupling constant $v(1) = v^*_{\rm
I}$.
\label{fig8}}

\figure{Fig.9: Effective exponent $\eta_{\perp \, {\rm eff}}(\ell)$ for the
pair correlation function as a function of the running anisotropy parameter
$g(\ell)$ for a fixed initial value of the coupling constant $v(1) = v^*_{\rm
I}$.
\label{fig9}}

\figure{Fig.10: Effective exponent $\gamma_{\rm eff}(\ell)$ for the
connectivity as a function of the running anisotropy parameter $g(\ell)$ for a
fixed initial value of the coupling constant $v(1) = v^*_{\rm I}$.
\label{fig10}}

\figure{Fig.11: Effective exponent $z_{\rm eff}(\ell)$, describing the
anisotropy, as a function of the running anisotropy parameter $g(\ell)$ for a
fixed initial value of the coupling constant $v(1) = v^*_{\rm I}$.
\label{fig11}}

\figure{Fig.12: Contour plots for $\Gamma_{11}(0,{\bf q},\omega)$, as inferred
from Eqs.~(\ref{4.62}) and (\ref{4.63}), with values of $\ell = 10^{-4}$,
$10^{-3}$, and $10^{-2}$, respectively.
\label{fig12}}


\begin{references}
\bibitem{Man82} B. B. Mandelbrot, {\em The Fractal Geometry of Nature},
                (Freeman, San Francisco, 1982);
                J. Feder, {\em Fractals}, (Plenum, New York, 1988).
\bibitem{Ess80} For a review on the theory of percolation, see: J. W. Essam,
                Rep. Prog. Phys. {\bf 43}, 834 (1980); D. Stauffer and A.
                Aharony, {\em Introduction to Percolation Theory}, 2nd ed.
                (Taylor and Francis, London, 1992).
\bibitem{Kin83} For a review, see W. Kinzel, in: {\em Percolation Structures
                and Processes}, Ann. Isr. Phys. Soc. Vol. 5, Eds. G. Deutscher,
                R. Zallen, and J. Adler (Bar-Ilan University, 1983), p. 425.
\bibitem{Car80} J. L. Cardy and R. L. Sugar, J. Phys. A {\bf 13}, L 423 (1980).
\bibitem{Bro78} R. Brower, M. A. Furman, and K. Subbarao, Phys. Lett.
                {\bf 78 B}, 213 (1978).
\bibitem{Hen90} M. Henkel and H. J. Herrmann, J. Phys. A {\bf 23}, 3719 (1990).
\bibitem{Gra89} P. Grassberger, J. Phys. A {\bf 22}, 3673 (1989).
\bibitem{Car84} J. Benzoni and J. L. Cardy, J. Phys. A {\bf 17}, 179 (1984).
\bibitem{Gra79} P. Grassberger and A. De La Torre, Ann. Phys. (N.Y.) {\bf 122},
                373 (1979).
\bibitem{Ben84} J. Benzoni, J. Phys. A {\bf 17}, 2651 (1984).
\bibitem{Obu80} S.P. Obukhov, Physica {\bf 101 A}, 145 (1980).
\bibitem{Lie81} N. Van Lieu and B. I. Shklovskii, Solid State Comm. {\bf 38},
                99 (1981).
\bibitem{Ker80} J. Kert\'esz and T. Vicsek, J. Phys. C {\bf 13}, L 343 (1980).
\bibitem{Gra85} P. Grassberger, J. Phys. A {\bf 18}, L 215 (1985).
\bibitem{Jan76} H. K. Janssen, Z. Phys. B {\bf 23}, 377 (1976);
                R. Bausch, H. K. Janssen, and H. Wagner, Z. Phys. B {\bf 24},
                113 (1976).
\bibitem{Ami78} D. J. Amit and Y. Y. Goldschmidt, Ann. Phys. (N.Y.) {\bf 114},
                356 (1978).
\bibitem{Ami84} D. J. Amit, {\em Field Theory, the Renormalization Group, and
                Critical Phenomena}, 2nd ed. (World Scientific, Singapore,
                1984).
\bibitem{Law81} I. D. Lawrie, J. Phys. A {\bf 14}, 2489 (1981);
                {\bf 18}, 1141 (1985).
\bibitem{Fre88} E. Frey and F. Schwabl, J. Phys. (Paris) Colloq. {\bf 48},
                C8-1569 (1988);
                Phys. Rev. B {\bf 42}, 8261 (1990);
                Phys. Rev. B {\bf 43}, 833 (1991).
\bibitem{Tau92} U. C. T\"auber and F. Schwabl, Phys. Rev. B {\bf 46}, 3337
                (1992).
\bibitem{Tau93} U. C. T\"auber and F. Schwabl, Phys. Rev. B {\bf 48}, 186
		(1993).
\bibitem{Hoo72} G. t'Hooft and M. Veltman, Nucl. Phys. B {\bf 44}, 189 (1972).
\bibitem{Eps}   However, in the usual $\epsilon$ expansion scheme, which
                constitutes a well--defined perturbation approach provided that
                $\epsilon \ll 1$, one finally has to insert values $\epsilon =
                1$, or even $\epsilon = 2$ or $3$ in the case of directed and
                isotropic percolation, respectively, in order to describe the
                cricital behavior in $d = 3$ dimensions.
\bibitem{Doh89} R. Schloms and V. Dohm, Nucl. Phys. B {\bf 328}, 639 (1989);
                Phys. Rev. B {\bf 42}, 6142 (1990).
\bibitem{Exp}   In most of the literature on directed percolation \cite{Car80}
                there is, unfortunately, a different definition for the
                critical exponents ${\tilde \eta}_\parallel$ and ${\tilde z}$.
                The relation to our definition is ${\tilde \eta}_\parallel = 1
                - \eta_\parallel $ and ${\tilde z} = z/2$. Our definition of
		the critical exponents corresponds to the usual convention used
		in the theory of static and dynamic critical phenomena.
\bibitem{Die86} H. W. Diehl, in: {\em Phase Transitions and Critical
                Phenomena}, Vol. 10, Eds. C. Domb and J. L. Lebowitz (Academic
                Press, London, 1986), p. 75.

\end{references}
\end{document}